# Comments on "On vector potential of the Coulomb gauge"*

A. M. Stewart

Emeritus Faculty,
The Australian National University,
Canberra, ACT 0200, Australia.

**Abstract**

The instantaneous nature of the potentials of the Coulomb gauge is clarified and a concise derivation is given of the vector potential of the Coulomb gauge expressed in terms of the instantaneous magnetic field.

We comment on a recent discussion [1, 4] of the speed of propagation of the potentials in the Coulomb gauge. In the Lorenz gauge, all components of the electromagnetic potential that originate from the charge and current sources propagate at the speed of light [3]. In the Coulomb gauge [3] the scalar potential acts instantaneously with respect to the charge density. However, from the expression for the vector potential obtained by Jackson [3]

$$A(x,t) = c\nabla_x \times \int d^3y \int_0^{R/c} d\tau\, \tau\, J(y, t-\tau) \times \hat{R}/R^2 \quad , \quad (1)$$

it can be seen that, with respect to the current source $J$, the vector potential propagates with different time delays, from $\tau = 0$, instantaneously, to $\tau = R/c$, where $R$ is the distance between $x$ and $y$, corresponding to the speed of light. If the vector potential of the Coulomb gauge is expressed in an equivalent form, in terms of the fields rather than the sources [2],

$$A(x,t) = \nabla_x \times \frac{1}{4\pi} \int d^3y\, \frac{B(y,t)}{|x-y|} \quad , \quad (2)$$

an expression that may be obtained either from the Helmholtz theorem [2] or by analogy with the Biot-Savart law, then the vector potential propagates instanteously with respect to the magnetic field. However this causes no conceptual difficulty because the fields themselves propagate at the speed of light. What counts is the velocity of propagation of the fields and this must always be the speed of light. The sole criterion for the validity of a given set of potentials is that they reproduce the correct fields; the velocity of propagation of the potentials is of no physical consequence.

To confirm that the vector potential given by equation (2) does indeed reproduce the correct fields it is necessary to show that it satisfies the equations

$$\boldsymbol{B} = \nabla \times \boldsymbol{A} \qquad \nabla \cdot \boldsymbol{A} = 0 \quad \text{with} \quad \nabla \cdot \boldsymbol{B} = 0 \quad . \tag{3}$$

The second of these conditions is satisfied by a standard vector identity. From the curl of (2) we get

$$\nabla_x \times \boldsymbol{A}(\boldsymbol{x},t) = \frac{1}{4\pi} \times \int d^3 y \, \nabla_x \times [\nabla_x \frac{1}{|\boldsymbol{x}-\boldsymbol{y}|} \times \boldsymbol{B}(\boldsymbol{y},t)] \quad , \tag{4}$$

and, by using a standard vector identity,

$$\nabla_x \times \boldsymbol{A}(\boldsymbol{x},t) = \frac{1}{4\pi} \times \int d^3 y \{-\nabla_x^2 \frac{1}{|\boldsymbol{x}-\boldsymbol{y}|} \boldsymbol{B}(\boldsymbol{y},t) + [\boldsymbol{B}(\boldsymbol{y},t) \cdot \nabla_x] \nabla_x \frac{1}{|\boldsymbol{x}-\boldsymbol{y}|} \} \quad . \tag{5}$$

The first term gives a delta function, leading to $\boldsymbol{B}(\boldsymbol{x},t)$. The second term becomes

$$-\frac{\nabla_x}{4\pi} \int d^3 y \, [\boldsymbol{B}(\boldsymbol{y},t) \cdot \nabla_y] \frac{1}{|\boldsymbol{x}-\boldsymbol{y}|} \quad , \tag{6}$$

and, using the vector identity for the divergence of the product of a vector and a scalar,

$$\nabla_y \cdot \frac{\boldsymbol{B}(\boldsymbol{y},t)}{|\boldsymbol{x}-\boldsymbol{y}|} = \boldsymbol{B}(\boldsymbol{y},t) \cdot \nabla_y \frac{1}{|\boldsymbol{x}-\boldsymbol{y}|} + \frac{\nabla_y \cdot \boldsymbol{B}(\boldsymbol{y},t)}{|\boldsymbol{x}-\boldsymbol{y}|} \quad , \tag{7}$$

becomes zero, because $\nabla \cdot \boldsymbol{B} = 0$ and the volume integral of the divergence gives rise to a surface integral that vanishes. Hence it is proved that for the potential of (2) $\boldsymbol{B}(\boldsymbol{x},t) = \nabla \times \boldsymbol{A}(\boldsymbol{x},t)$.

Eq. (2) has been used to obtain an expression for the bound angular momentum (that part associated with a charge density) of the electromagnetic field [5] and to show that the electromagnetic field makes zero contribution to the angular momentum of the physical electron described by the Lagrangian of quantum electrodynamics [6].